# PRELIMINARY STUDY OF MORTALITY BY CAUSE AND SOCIODEMOGRAPHIC CHARACTERISTICS. MUNICIPALITY OF SAN FRANCISCO - ANTIOQUIA, 2001-2010.

## ESTUDIO PRELIMINAR DE LA MORTALIDAD POR CAUSA Y CARACTERÍSTICAS SOCIODEMOGRÁFICAS. MUNICIPIO DE SAN FRANCISCO - ANTIOQUIA, 2001-2010.

## Working Paper

*Dayana L. Jiménez V[1]; Paola A. Gutiérrez L[2]; Yeisson A. Gutiérrez C[3]; Fernán A. Villa G[4]*

1. Gerencia de Sistemas de Información en Salud, Facultad Nacional de Salud Pública, Universidad de Antioquia, Medellín, Colombia. Correo electrónico: dayajimenes@gmail.com
2. Gerencia de Sistemas de Información en Salud, Facultad Nacional de Salud Pública, Universidad de Antioquia, Medellín, Colombia. Correo electrónico: pao.gutierrez528@gmail.com
3. Gerencia de Sistemas de Información en Salud, Facultad Nacional de Salud Pública, Universidad de Antioquia, Medellín, Colombia. Correo electrónico: ygutierrezcano@gmail.com
4. Doctor en Ingeniería – Sistemas e informática, Universidad Nacional de Colombia, Facultad de Minas, Medellín, Colombia. Correo electrónico: favillao@unal.edu.co

**Resumen**

**Objetivo:** Determinar la estructura de la mortalidad registrada por causa y las características sociodemográficas, en el municipio de San Francisco - Antioquia, 2001-2010. **Metodología:** Estudio cuantitativo de tipo descriptivo con información longitudinal retrospectiva obtenida de fuente secundaria de los eventos de defunción a través de las bases de datos en medio electrónico suministradas por el DANE. Se realizó una descripción de las variables sociodemográficas según grupos de causa de mortalidad, se calculó la tabla de vida por sexo y los años potenciales de vida perdidos (APVP) para cada año y grupo de causa de muerte. **Resultados:** Se presentó las causas externas y las agresiones, homicidios, como principal causa de muerte durante el decenio de estudio, y principalmente en los hombres, los cuales presentaron mayores tasas de mortalidad, mayor probabilidad de morir y menos esperanza de vida durante el periodo. En promedio los hombres y las causas externas presentaron mayor número de años potenciales de vida perdidos para los años 2001-2010. **Conclusiones:** Los hombres presentan mayores tasas de mortalidad durante el decenio, al igual que las causas externas, agresiones, homicidios, y en mayor proporción en los hombres jóvenes. Las causas de muerte que aportan mayor número de años potenciales de vida perdidos son las causas externas. La esperanza de vida al nacer y durante todo el decenio es mayor para las mujeres que para los hombres, por ello, es indispensable que el municipio analice la situación actual de estas causas de muerte, ya que deben llevarse a cabo políticas públicas que aporten a su disminución y por ende a mejorar la esperanza de vida de la población.

**Palabras clave:** Defunción, Esperanza de vida, APVP, Mortalidad.

**Abstract**

**Objective:** Determining the structure of mortality from causes and sociodemographic characteristics, in the municipality of San Francisco - Antioquia, 2001-2010. **Methodology:** Quantitative descriptive study with retrospective longitudinal data obtained from secondary source of death events through databases in electronic media supplied by the DANE. A description of the sociodemographic variables was performed by groups of cause of death, the life table by sex and years of potential life lost (APVP) for each year group and cause of death was calculated. **Results:** External causes and assaults, homicides, as the main cause of death occurs during the decade of study, and especially in men, which had higher mortality rates, more likely to die and less life expectancy during the period. On average men and external causes showed a higher number of potential years of life lost for the years 2001-2010 life. **Conclusions:** Men have higher mortality rates over the decade, as external causes, assaults, murders, and a greater proportion of young men. The causes of death that bring more potential years of life lost are external causes. Life expectancy at birth and throughout the decade is greater for women than it is therefore essential for men, that the municipality analyze the current situation of these causes of death, and to be carried out public policies that contribute to its decline and hence to improve the life expectancy of the population.

**Keywords:** Death, Lifespan, APVP, Mortality.

**Introducción**

"La muerte es la desaparición permanente de todo signo de vida cualquiera que sea el tiempo transcurrido desde el nacimiento vivo", y al referirse a todas las muertes que son notificadas en una población se habla de mortalidad, siendo esta uno de los determinantes de su crecimiento natural. La mortalidad permite medir el nivel de vida, la estructura de salud e identificar los problemas más relevantes en el tiempo y en el espacio. La especie humana está destinada a morir algún día, la probabilidad de morir durante un periodo determinado del tiempo se relaciona con muchos factores sociodemográficos y se presenta por diferentes causas, ya sean endógenas o exógenas, las cuales varían según la población y el tiempo, y se ven influenciadas por diferentes factores ambientales y de salud (1-5).

Para el registro y notificación de las defunciones se utiliza el certificado de defunción, éste es fuente primaria de información, con los datos que este aporta se construyen indicadores básicos de mortalidad, los cuales describen el comportamiento de este fenómeno en la población documentada. Igualmente, la mortalidad se puede analizar teniendo en cuenta los eventos que son causa de ella, como lo son las enfermedades, estados morbosos o lesiones que causaron la muerte o que contribuyeron a ella, y las circunstancias del accidente o de la violencia que produjeron dichas lesiones. Es significativo resaltar, la variable sociodemográfica sexo, puesto que aparte de existir diferencia en lo biológico y en lo social, se presenta también en la mortalidad, como consecuencia de ello en la esperanza de vida; para la Organización Mundial de la Salud (OMS) la diferencia de la esperanza de vida entre las mujeres y los hombres ha tenido una tendencia creciente en los últimos años. Por lo cual se hace interesante el análisis de la mortalidad por causa, sexo y demás factores involucrados en ella. (3,6-8)

La mortalidad se expresar en términos de números absolutos, tasas y proporciones. No basta solo con saber que la esperanza de vida está en aumento, es más relevante si se indica su magnitud y su distribución por medio de datos accesibles a los investigadores y comprensibles para los lectores interesados en la información; expresados en número absolutos, tasas o razones. Una tasa expresa la frecuencia de los eventos ocurridos en una población durante un periodo determinado de tiempo, que normalmente es un año; es decir, cuán común es que suceda este evento. La razón se expresa como la relación que existe entre un subgrupo de la población y la población total u otro subgrupo; es decir, un subgrupo dividido por otro. La proporción se refiere al tamaño de la ocurrencia de un evento con respecto al total de la población (5).

La tabla de vida es un elemento importante para aportar información sobre la mortalidad en un periodo determinado; esta es una matriz en la que se muestra las probabilidades de vida y de muerte en una población, al igual que la esperanza de vida, la cual toma como base una generación hipotética de la población desde su nacimiento y sigue a través de toda su vida hasta su total extinción. De igual forma el análisis de la mortalidad puede ser abordada desde otro punto de vista, tratando de medir el tiempo que se deja de vivir. Los Años Potenciales de Vida Perdidos (APVP) son un conjunto de medidas que se fundamentan bajo el concepto de mortalidad prematura, dónde cada muerte es valorada contemplando los años que hipotéticamente ha dejado de vivir la persona que fallece, es decir, cuanto tiempo en años de vida se pierde por cada muerte ocurrida. Generalmente se utiliza la esperanza de vida como límite superior para calcular este indicador, y su utilización es muy importante, ya que son especialmente apropiados para medir el impacto de las causas de muerte en una población. Además, son útiles para establecer prioridades de intervención o de investigación en los sistemas de salud, orientar la dotación de recursos o comparar la efectividad de distintas actuaciones (9,10).

Estudiar el comportamiento de la mortalidad a nivel mundial, nacional, departamental y municipal es una base fundamental para la toma de decisiones en aspectos de salud pública dentro de una población (1). Basado en ello, se encontró que actualmente, de los 56 millones de muertes que ocurren en el mundo al año, el número de muertes por enfermedades cardiovasculares es de 7.3 millones (13%), por infecciones de vías respiratorias y pulmonares se presentan más de 6.7 millones (11.9%) de muertes al año, al igual que por enfermedades cerebrovasculares que es de 6.15 millones (10.9%), por enfermedades diarreicas se presentan 2.5 millones de muertes al año (4.5%); el sida mata cada año cerca de 2 millones de personas (3.6%) y el cáncer de pulmón y de tráquea a cerca de 1.4 millones de personas al año (2.5%) (11).

Cada año en américa latina, se producen cerca de 3,570,000 muertes y es una de las regiones que a nivel mundial tiene la tasa más baja de mortalidad. Para la Organización Panamericana de la Salud OPS, la esperanza de vida aumentó de 66 a 72 años desde 1980, mientras que al mismo tiempo la mortalidad en el mundo ha disminuido casi en un 25%. Las infecciones respiratorias agudas y las diarreas todavía afectan a muchos niños, especialmente en las zonas rurales y los suburbios urbanos que representan el 75% de la población (12).

En Colombia, para el periodo de 1985 al 2006, en los hombres la mayor proporción de defunciones corresponde a las causas externa (20%), seguido de las enfermedades del sistema circulatorio con el 15% y los tumores con el 7%. Entre las mujeres, el primer lugar lo ocupan las enfermedades del sistema circulatorio con un 14%, seguidas por los tumores con un 7.4%, las causas de enfermedades respiratorias con un 4% y las causas externas con un 3.1% (13). También, para el departamento de Antioquia en el año 2009 se encontraron como principales causas de muerte las agresiones (homicidios) con 4 975 muertes (16.4%), las enfermedades isquémicas del corazón con 4 223 defunciones (13.9%) y las enfermedades crónicas respiratorias con 1 979 muertes (6.5%) (14). El municipio de San Francisco - Antioquia, para el año 2010 posee una población de 5,838 habitantes en sus diferentes veredas y cabecera municipal (15). Según el Departamento Administrativo Nacional de Estadística (DANE) para los años de 2001 a 2010 se presentaron en promedio 43 muertes por año. Las causas de muerte con mayor frecuencia en estos mismos años han sido las enfermedades isquémicas del corazón con un promedio del 17.9% y las agresiones u homicidios con un 15.4% en promedio (16).

Tras la búsqueda en las bases de datos bibliográficas a nivel nacional, publicaciones del DANE y la página de la administración del municipio, se evidenció que no se dispone un estudio sobre la mortalidad por causa y la relación de ésta con las características sociodemográficas de la población para los años de 2001 – 2010 del municipio de San Francisco - Antioquia.

En el presente estudio, se propuso describir la estructura de la mortalidad registrada por causa y las características sociodemográficas, en el municipio de San Francisco - Antioquia, para los años comprendidos entre 2001-2010, con el fin de nutrir de nueva información sobre mortalidad por causa con respecto a las características sociodemográficas al sector salud y a los entes gubernamentales para la toma de decisiones en relación con los programas de promoción de la salud y prevención de la enfermedad en la población de dicho municipio.

**Metodología**

Se realizó un estudio cuantitativo de tipo descriptivo con información longitudinal retrospectiva, que se obtuvo de fuente secundaria a través de las bases de datos en medio electrónico de las defunciones registradas en Colombia entre 1980-2010, la cual fue suministrada por la Facultad Nacional de Salud Pública de la Universidad de Antioquia, homologada y ajustada por la Línea Análisis de la Situación de Salud del grupo de Investigación Epidemiología, teniendo como fuente de información básica el DANE; para determinar la estructura de la mortalidad registrada por causa y las características sociodemográficas, en el municipio de San Francisco - Antioquia, 2001-2010.

La base de datos contaba con 395 registros del componente de defunción de las estadísticas vitales del DANE de las personas residentes del municipio de San Francisco – Antioquia para los años 2001 – 2010, los cuales fueron seleccionados en su totalidad para la realización del estudio (17).

Se realizó la descripción de las defunciones para los 7 grandes grupos de causas agrupadas con base en la lista 6/67 de la Organización Panamericana de la Salud OPS (CIE-10) según características sociodemográficas, por medio de una tabla que muestra las proporciones y las tasas de mortalidad para cada una de ellas. Para la descripción de las variables sociodemográficas y las principales causas de muerte se construyeron tablas de frecuencia con las proporciones y tasas, teniendo en cuenta las 5 primeras causas de muerte agrupadas con base en la lista 6/67 de la (OPS) (CIE-10); se realizaron gráficos para la descripción de las tasas de mortalidad para cada sexo según grupos quinquenales de edad, de igual forma para las primeras cinco causas de muerte.

Se construyó gráficos para la descripción de la esperanza de vida para cada sexo en los grupos quinquenales de edad, y para estas mismas variables se realizó el grafico de la probabilidad de morir; datos extraídos de la construcción de las tablas de vida para cada sexo durante el decenio de estudio.

Para la descripción de los años potenciales de vida perdidos (APVP) se calcularon por año para cada sexo, y de igual forma fueron calculados según los 7 grandes grupos de causas agrupadas para todo el decenio.

Los cálculos de las tablas de frecuencia para la descripción de las defunciones según variables sociodemográficas y causa de muerte, y para la construcción de los gráficos de las tasas de mortalidad, se realizó por medio del software estadístico SPSSS 20. El cálculo de las tablas de vida, de las cuales se construyó los gráficos de esperanza de vida y probabilidad de morir, y el cálculo de los APVP por grupo de causas de muerte, se realizó mediante el software estadístico Epidat 3.1.

**Resultados**

En el análisis de los datos se evidenciaron algunos registros sin información de algunas variables, estos no fueron incluidos en los resultados por falta de relevancia frente a los mismos.

Del 2001 al 2010, San Francisco presentó una mayor proporción de defunciones en los hombres con un 75.7% (863 defunciones por cada 100,000 hombres) que en las mujeres con un 24.3% (306 muertes por cada 100,000 mujeres), la tasa bruta de mortalidad fue de 623 defunciones por cada 100,000 habitantes para este decenio. Por estado civil se encontró que la proporción para las personas que estaban casadas fue de 29.6% (185 muertes por cada 100,000 habitantes) y para las personas separadas o divorciadas fue de 0.5% (3 muertes por cada 100,000 habitantes) las cuales presentaron el más bajo porcentaje y tasa de mortalidad durante este periodo. (Tabla 1)

Para describir las defunciones según el nivel educativo, se encontró que San Francisco presentó este fenómeno en mayor proporción en las personas con un nivel educativo de primaria con un 28.1% (175 defunciones por cada 100,000 habitantes); las personas que registraron la menor proporción de defunciones en este periodo fueron las que tenían un nivel educativo de preescolar con un 0.3% (3 muertes por cada 100,000 habitantes). Las personas del régimen subsidiado presentaron una mayor proporción de defunciones con un 48.1% (299 defunciones por cada 100,000 habitantes) en comparación con los del régimen contributivo que fue de 5.3% (33 muertes por cada 100,000 habitantes) y otro tipo de afiliación con un 15.2% (95 defunciones por cada 100,000 habitantes). (Tabla 1)

La mayor proporción de defunciones se presentó en el área rural disperso con un 56.7% (353 muertes por cada 100,000 habitantes), seguido de la cabecera municipal con un 35.2% (219 defunciones por cada 100,000 habitantes) y el centro poblado con un 6.1% (38 muertes por cada 100,000 habitantes). (Tabla 1)

Al observar las defunciones entre los 7 grandes grupos de causas agrupadas con base en la lista 6/67 de la OPS (CIE-10) (Tabla 1), se encontró un predominio de muertes por causas externas; para los hombres la proporción de muertes por causas externas fue de 73.6% (351 muertes por cada 100,000 habitantes), mayor a la de las mujeres que fue de 21.9% (36 muertes por cada 100,000 habitantes). La enfermedades del sistema circulatorio fue la única causa de muerte presentada entre las personas separadas con 3 muertes por cada 100,000 habitantes, las causas externas igualmente fue la única causa de muerte entre las personas con un nivel educativo de preescolar y secundaria, 2 muertes por cada 100,000 habitantes y 22 defunciones por cada 100,000 habitantes respectivamente. (Tabla 1)

Tabla 1. Proporciones y Tasas de mortalidad según causa de defunción agrupada en los 7 grandes grupos con base en la lista 6/67 de la OPS (CIE-10), y las características sociodemográficas, municipio de San Francisco – Antioquia, decenio 2001-2010.

| | Causa agrupada con base en la Lista 6/67 de la OPS (CIE-10) | | | | | | | | | | | | | | |
|---|---|---|---|---|---|---|---|---|---|---|---|---|---|---|---|
| | Enfermedades transmisibles | | Neoplasias (tumores) | | Enfermedades del sistema circulatorio | | Ciertas afecciones originado en el periodo perinatal | | Causas externas | | Todas las demás causas | | Síntomas, signos y afecciones mal definidas | |
| | n (%) | Tasa x 100,000 | n (%) | Tasa x 100,000 | n (%) | Tasa x 100,000 | n (%) | Tasa x 100,000 | n (%) | Tasa x 100,000 | n (%) | Tasa x 100,000 | n (%) | Tasa x 100,000 |
| **Área** | | | | | | | | | | | | | | |
| Cabecera municipal | 5 (3.5) | 8.0 | 15 (10.6) | 23.9 | 34 (24.1) | 54.2 | 1 (0.7) | 1.6 | 58 (41.1) | 92.5 | 16 (11.3) | 25.5 | 12 (8.5) | 19.1 |
| Centro poblado | 2 (6.8) | 3.2 | 1 (3.4) | 1.6 | 4 (13.7) | 6.4 | 0 - | 0.0 | 19 (65.5) | 30.3 | 1 (3.4) | 1.6 | 2 (6.9) | 3.2 |
| Rural disperso | 3 (1.8) | 4.8 | 11 (6.7) | 17.5 | 13 (8.0) | 20.7 | 4 (2.5) | 6.4 | 104 (64.2) | 165.9 | 11 (6.8) | 17.5 | 16 (9.9) | 25.5 |
| **Sexo** | | | | | | | | | | | | | | |
| Hombres | 5 (1.6) | 8.0 | 15 (5.0) | 23.9 | 30 (10.0) | 47.8 | 2 (0.7) | 3.2 | 220 (73.6) | 350.9 | 11 (3.7) | 17.5 | 16 (5.4) | 25.5 |
| Mujeres | 5 (5.2) | 8.0 | 13 (13.5) | 20.7 | 22 (22.9) | 35.1 | 3 (3.1) | 4.8 | 21 (21.9) | 33.5 | 17 (17.7) | 27.1 | 15 (15.6) | 23.9 |
| **Estado civil** | | | | | | | | | | | | | | |
| Unión libre | 0 - | 0.0 | 1 (4.5) | 1.6 | 4 (18.2) | 6.4 | 0 - | 0.0 | 16 (72.7) | 25.5 | 1 (4.5) | 1.6 | 0 - | 0.0 |
| Separado | 0 - | 0.0 | 0 - | 0.0 | 2 (100) | 3.2 | 0 - | 0.0 | 0 - | 0.0 | 0 - | 0.0 | 0 - | 0.0 |
| Viudo | 1 (2.4) | 1.6 | 8 (19.5) | 12.8 | 17 (41.5) | 27.1 | 0 - | 0.0 | 2 (4.9) | 3.2 | 7 (17.0) | 11.2 | 6 (14.6) | 9.6 |
| Soltero | 3 (3.1) | 4.8 | 2 (2.1) | 3.2 | 7 (7.3) | 11.2 | 5 (5.2) | 8.0 | 64 (66.7) | 102.1 | 9 (9.4) | 14.4 | 6 (6.3) | 9.6 |
| Casado | 5 (4.2) | 8.0 | 15 (12.8) | 23.9 | 21 (17.9) | 33.5 | 0 - | 0.0 | 47 (40.2) | 75.0 | 11 (9.4) | 17.5 | 18 (15.4) | 28.7 |
| **Nivel educativo** | | | | | | | | | | | | | | |
| Preescolar | 0 - | 0.0 | 0 - | 0.0 | 0 - | 0.0 | 0 - | 0.0 | 1 (100) | 1.6 | 0 - | 0.0 | 0 - | 0.0 |
| Primaria | 3 (2.7) | 4.8 | 17 (15.3) | 27.1 | 14 (12.6) | 22.3 | 0 - | 0.0 | 60 (54.1) | 95.7 | 9 (8.1) | 14.4 | 8 (7.2) | 12.8 |
| Secundaria | 0 - | 0.0 | 0 - | 0.0 | 0 - | 0.0 | 0 - | 0.0 | 14 (100) | 22.3 | 0 - | 0.0 | 0 - | 0.0 |
| Educación superior | 0 - | 0.0 | 0 - | 0.0 | 0 - | 0.0 | 0 - | 0.0 | 0 - | 0.0 | 0 - | 0.0 | 0 - | 0.0 |
| Ninguno | 2 (2.7) | 3.2 | 6 (8.3) | 9.6 | 17 (23.6) | 27.1 | 0 - | 0.0 | 21 (29.2) | 33.5 | 14 (19.4) | 22.3 | 12 (16.7) | 19.1 |
| **Régimen** | | | | | | | | | | | | | | |
| Contributivo | 0 - | 0.0 | 3 (14.2) | 4.8 | 2 (9.5) | 3.2 | 0 - | 0.0 | 14 (66.7) | 22.3 | 1 (4.8) | 1.6 | 1 (4.8) | 1.6 |
| Subsidiado | 9 (4.7) | 14.4 | 23 (12.1) | 36.7 | 43 (22.6) | 68.6 | 4 (2.1) | 6.4 | 63 (33.2) | 100.5 | 22 (11.6) | 35.1 | 26 (13.7) | 41.5 |
| Otros | 0 - | 0.0 | 0 - | 0.0 | 4 (6.7) | 6.4 | 0 - | 0.0 | 52 (86.7) | 82.9 | 1 (1.7) | 1.6 | 3 (5.0) | 4.8 |

La proporción de muertes por agresiones (homicidios), que fue la predominante durante el decenio, en hombres fue de un 94% (249 muertes por cada 100,000 habitantes) siendo mucho mayor en comparación con las mujeres que solo fue de un 6% (16 muertes por cada 100,000 habitantes) (Tabla 2). De igual forma esta proporción fue mayor en las personas solteras con un 29.8% (79 muertes por cada 100,000 habitantes) y en las personas con un nivel educativo de primaria 28.6% (76 defunciones por cada 100,000 habitantes). (Tabla 2)

Tabla 2. Proporciones y Tasas la mortalidad según causa de defunción agrupada con base en la lista 6/67 de la OPS (CIE-10), con respecto a las características sociodemográficas, municipio de San Francisco – Antioquia, decenio 2001-2010.

| | Causa de defunción agrupada con base en la lista 6/67 de la OPS (CIE-10) | | | | | | | | | |
|---|---|---|---|---|---|---|---|---|---|---|
| | Enfermedades isquémicas del corazón | | Enfermedades cerebrovasculares | | Agresiones (homicidios), inclusive secuelas | | Intervención legal y operaciones de guerra, inclusive secuelas | | Enfermedades crónicas de las vías respiratorias inferiores | |
| | n (%) | Tasa x 100,000 | n (%) | Tasa x 100,000 | n (%) | Tasa x 100,000 | n (%) | Tasa x 100,000 | n (%) | Tasa x 100,000 |
| **Área** | | | | | | | | | | |
| Cabecera municipal | 23 (79.3) | 36.3 | 12 (80.0) | 18.9 | 21 (12.5) | 33.1 | 4 (7.7) | 6.3 | 6 (54.5) | 9.5 |
| Centro poblado | 2 (6.9) | 3.2 | 1 (6.7) | 1.6 | 12 (7.1) | 18.9 | 1 (1.9) | 1.6 | 1 (9.1) | 1.6 |
| Rural disperso | 3 (10.3) | 4.7 | 2 (13.3) | 3.2 | 130 (77.4) | 205.0 | 46 (88.5) | 72.5 | 4 (36.4) | 6.3 |
| **Sexo** | | | | | | | | | | |
| Hombres | 17 (58.6) | 26.8 | 8 (53.3) | 12.6 | 158 (94.0) | 249.2 | 44 (84.6) | 69.4 | 8 (72.7) | 12.6 |
| Mujeres | 12 (41.4) | 18.9 | 7 (46.7) | 11.0 | 10 (6.0) | 15.8 | 8 (15.4) | 12.6 | 3 (27.3) | 4.7 |
| **Estado civil** | | | | | | | | | | |
| Unión libre | 3 (10.3) | 4.7 | 1 (6.7) | 1.6 | 14 (8.3) | 22.1 | 2 (3.8) | 3.2 | 1 (9.1) | 1.6 |
| Separado | 1 (3.4) | 1.6 | - | - | - | - | - | - | - | - |
| Viudo | 11 (37.9) | 17.3 | 4 (26.7) | 6.3 | 2 (1.2) | 3.2 | - | - | 3 (27.3) | 4.7 |
| Soltero | 3 (10.3) | 4.7 | 3 (20.0) | 4.7 | 50 (29.8) | 78.9 | 8 (15.4) | 12.6 | 5 (45.5) | 7.9 |
| Casado | 10 (34.5) | 15.8 | 7 (46.7) | 11.0 | 38 (22.6) | 59.9 | 1 (1.9) | 1.6 | 2 (18.2) | 3.2 |
| **Nivel educativo** | | | | | | | | | | |
| Preescolar | - | - | - | - | 1 (0.6) | - | - | - | - | - |
| Primaria | 11 (37.9) | 17.3 | 2 (13.3) | 3.2 | 48 (28.6) | 75.7 | 7 (13.5) | 11.0 | 3 (27.3) | 4.7 |
| Secundaria | - | - | - | - | 10 (6.0) | 119.4 | 2 (3.8) | 17.4 | - | - |
| Educación superior | - | - | - | - | - | - | - | - | - | - |
| Ninguno | 9 (31.0) | 14.2 | 5 (33.3) | 7.9 | 16 (9.5) | 25.2 | 2 (3.8) | 3.2 | 7 (63.6) | 11.0 |
| **Régimen** | | | | | | | | | | |
| Contributivo | 1 (3.4) | 1.6 | 1 (6.7) | 1.6 | 11 (6.5) | 17.3 | 1 (1.9) | 1.6 | - | - |
| Subsidiado | 25 (86.2) | 39.4 | 11 (73.3) | 17.3 | 48 (28.6) | 75.7 | 5 (9.6) | 7.9 | 10 (90.9) | 15.8 |
| Otros | 1 (3.4) | 1.6 | 3 (20.0) | 4.7 | 32 (19.0) | 50.5 | 14 (26.9) | 22.1 | - | - |

Para las primeras 5 causas de defunción agrupadas con base en la lista 6/67 de la OPS (CIE-10), se presentó mayor proporción para el grupo de agresiones, homicidios, inclusive secuelas con un 61.1% (168) con predominio en la población joven; y el grupo de enfermedades crónicas de las vías respiratorias inferiores se presentó en menor proporción con un 4% (11) siendo esta predominante en la población adulta. En los grupos de enfermedades isquémicas del corazón y enfermedades cerebrovasculares se observó que empiezan a presentarse después de los 45 años con una proporción de 10.5% (29) y 5.5% (15) respectivamente; el grupo de causa de defunción que también presentó predominio en la población joven fue intervención legal y operaciones de guerra, inclusive secuelas con una proporción de 18.9% (52). (Tabla 2)

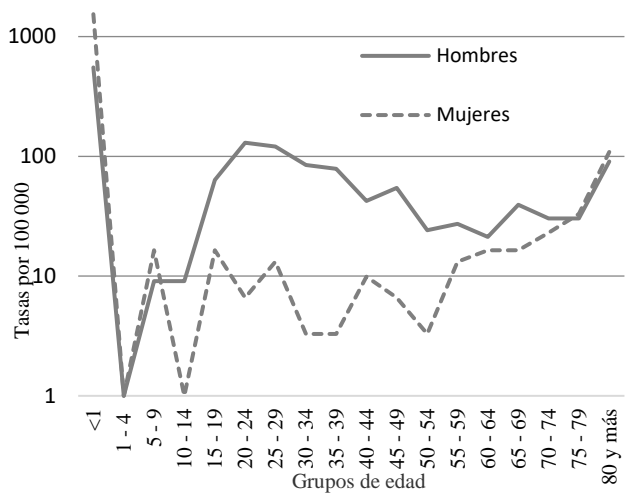

Figura 1. Tasa de mortalidad por 100,000 habitantes según sexo y grupo de edad, San Francisco – Antioquia, 2001-2010

Se presentó una gran diferencia de la tasa de mortalidad por sexo entre los 10 y los 54 años de edad, siendo ésta mucho mayor en los hombres (Figura 1), con un pico de la línea de tendencia en el grupo de edad 20-24 años, indicando una mayor mortalidad para ellos en este grupo de edad. (Figura 1)

Al hacer una comparación entre las tasas de mortalidad para cada grupo de causas de defunción, se observó de manera reiterativa, el predominio del número de muertes causadas por las agresiones, homicidios, inclusive secuelas durante todo el decenio. El grupo de causas de defunción que presentó menor tasa de mortalidad fue las enfermedades crónicas de las vías respiratorias inferiores. (Figura 2)

Se observó que comparando el sexo según el grupo de causas de mortalidad, en los hombres predomina el número de muertes en todos los grupos, en comparación con las mujeres que registran menor número de defunciones durante todo el decenio (Figura 2). Al realizar una razón por cada causa de muerte para el sexo se encontró que, por cada mujer que murió por agresión, homicidio, inclusive secuelas murieron aproximadamente 15 hombre (H/M = 14.5), por cada mujer que murió a causa de intervención legal y operaciones de guerra, inclusive secuelas murieron aproximadamente 5 hombres (H/M = 5.1) y por cada mujer que murió a causa de enfermedades crónicas de la vías respiratoria inferiores murieron aproximadamente 3 hombres (H/M = 2.5). (Figura 2).

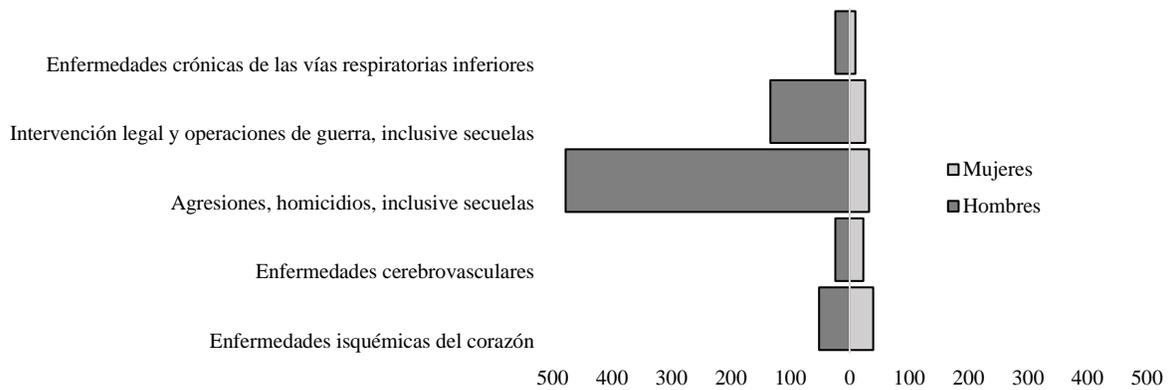

Figura 2. Tasa de mortalidad por 100,000 habitantes según causa de defunción agrupada con base en la lista 6/67 de la OPS (CIE-10) para cada sexo, San Francisco – Antioquia, 2001-2010.

En promedio las mujeres presentaron una mayor esperanza de vida al nacer durante el decenio (86 años), en comparación con los hombres (67.7 años) que su esperanza de vida al nacer fue de 20 años menos aproximadamente; este predominio de la esperanza de vida en las mujeres se mantuvo durante todos los grupos de edad hasta los 59 años en donde se empezó a presentar un esperanza de vida igual para ambos sexos. (Figura 3)

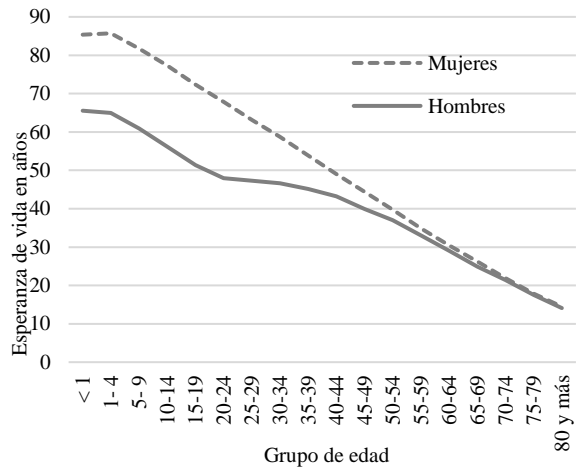

Figura 3. Esperanza de vida por grupo de edad para cada sexo, San Francisco – Antioquia, 2001-2010.

Al hacer un análisis de la probabilidad de morir por sexo para todo el decenio, se registra durante el transcurso del mismo un predominio de los hombres con respecto a las mujeres, las cuales presentaron menor probabilidad de morir durante este periodo, excepto al nacer que la probabilidad de las mujeres es mayor. (Figura 4)

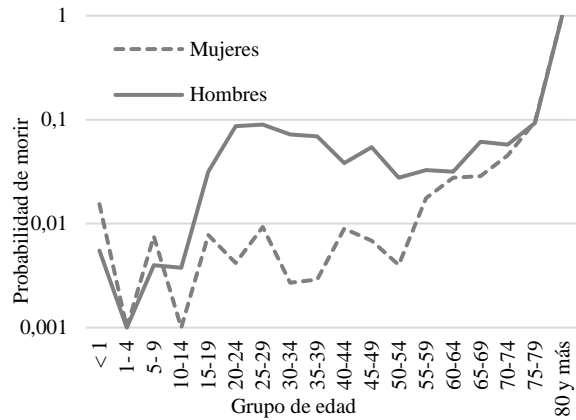

Figura 4. Probabilidad de morir por grupo de edad para cada sexo, San Francisco – Antioquia, 2001-2010.

Las tablas de vida para cada sexo fueron las siguientes, hombres y mujeres respectivamente, en promedio para los años 2001-2010. (Tabla 3 y 4)

Tabla 3. Tabla de vida para los hombres, municipio de San Francisco – Antioquia, decenio 2001-2010.

| Grupo | Población | $d_x$ | $_nm_x$ | $_nq_x$ | $_np_x$ | $l_x$ | $_nd_x$ | $_nL_x$ | $T_x$ | $e_x$ |
|---|---|---|---|---|---|---|---|---|---|---|
| < 1 | 362 | 2 | 0.00555 | 0.00552 | 0.99448 | 100000 | 552 | 99480 | 6555374 | 65.55 |
| 1- 4 | 2990 | 0 | 0 | 0 | 1 | 99448 | 0 | 397790 | 6455894 | 64.92 |
| 5- 9 | 3755 | 3 | 0.0008 | 0.00399 | 0.99601 | 99448 | 396 | 496246 | 6058104 | 60.92 |
| 10-14 | 3979 | 3 | 0.00075 | 0.00376 | 0.99624 | 99051 | 373 | 494323 | 5561858 | 56.15 |
| 15-19 | 3278 | 21 | 0.00641 | 0.03153 | 0.96847 | 98678 | 3111 | 485614 | 5067534 | 51.35 |
| 20-24 | 2374 | 43 | 0.01811 | 0.08664 | 0.91336 | 95567 | 8280 | 457137 | 4581920 | 47.94 |
| 25-29 | 2129 | 40 | 0.01879 | 0.08973 | 0.91027 | 87287 | 7832 | 416856 | 4124784 | 47.26 |
| 30-34 | 1867 | 28 | 0.015 | 0.07228 | 0.92772 | 79455 | 5743 | 382920 | 3707927 | 46.67 |
| 35-39 | 1814 | 26 | 0.01433 | 0.06919 | 0.93081 | 73713 | 5100 | 355813 | 3325008 | 45.11 |
| 40-44 | 1793 | 14 | 0.00781 | 0.03829 | 0.96171 | 68613 | 2627 | 336495 | 2969195 | 43.27 |
| 45-49 | 1609 | 18 | 0.01119 | 0.05441 | 0.94559 | 65985 | 3590 | 320950 | 2632700 | 39.9 |
| 50-54 | 1428 | 8 | 0.0056 | 0.02762 | 0.97238 | 62395 | 1724 | 307665 | 2311750 | 37.05 |
| 55-59 | 1347 | 9 | 0.00668 | 0.03286 | 0.96714 | 60671 | 1994 | 298372 | 2004085 | 33.03 |
| 60-64 | 1089 | 7 | 0.00643 | 0.03163 | 0.96837 | 58678 | 1856 | 288748 | 1705713 | 29.07 |
| 65-69 | 1030 | 13 | 0.01262 | 0.06118 | 0.93882 | 56822 | 3476 | 275417 | 1416965 | 24.94 |
| 70-74 | 843 | 10 | 0.01186 | 0.0576 | 0.9424 | 53345 | 3073 | 259045 | 1141547 | 21.4 |
| 75-79 | 512 | 10 | 0.01953 | 0.09311 | 0.90689 | 50273 | 4681 | 239660 | 882503 | 17.55 |
| 80 y más | 423 | 30 | 0.07092 | 1 | 0 | 45592 | 45592 | 642842 | 642842 | 14.1 |

Tabla 4. Tabla de vida para las mujeres, municipio de San Francisco – Antioquia, decenio 2001-2010.

| Grupo | Población | $d_x$ | $_nm_x$ | $_nq_x$ | $_np_x$ | $l_x$ | $_nd_x$ | $_nL_x$ | $T_x$ | $e_x$ |
|---|---|---|---|---|---|---|---|---|---|---|
| < 1 | 325 | 5 | 0.0156 | 0.01538 | 0.98462 | 100000 | 1538 | 98609 | 8537509 | 85.38 |
| 1- 4 | 2488 | 0 | 0 | 0 | 1 | 98462 | 0 | 393846 | 8438900 | 85.71 |
| 5- 9 | 3290 | 5 | 0.00152 | 0.00757 | 0.99243 | 98462 | 745 | 490444 | 8045054 | 81.71 |
| 10-14 | 3689 | 0 | 0 | 0 | 1 | 97716 | 0 | 488581 | 7554609 | 77.31 |
| 15-19 | 3197 | 5 | 0.00156 | 0.00779 | 0.99221 | 97716 | 761 | 486678 | 7066029 | 72.31 |
| 20-24 | 2389 | 2 | 0.00084 | 0.00418 | 0.99582 | 96955 | 405 | 483763 | 6579350 | 67.86 |
| 25-29 | 2145 | 4 | 0.00186 | 0.00928 | 0.99072 | 96550 | 896 | 480510 | 6095588 | 63.13 |
| 30-34 | 1851 | 1 | 0.00054 | 0.0027 | 0.9973 | 95654 | 258 | 477625 | 5615078 | 58.7 |
| 35-39 | 1722 | 1 | 0.00058 | 0.0029 | 0.9971 | 95396 | 277 | 476288 | 5137453 | 53.85 |
| 40-44 | 1665 | 3 | 0.0018 | 0.00897 | 0.99103 | 95119 | 853 | 473464 | 4661165 | 49 |

| Grupo | Población | $d_x$ | $_nm_x$ | $_nq_x$ | $_np_x$ | $l_x$ | $_nd_x$ | $_nL_x$ | $T_x$ | $e_x$ |
|---|---|---|---|---|---|---|---|---|---|---|
| 45-49 | 1460 | 2 | 0.00137 | 0.00683 | 0.99317 | 94266 | 643 | 469723 | 4187701 | 44.42 |
| 50-54 | 1243 | 1 | 0.0008 | 0.00401 | 0.99599 | 93623 | 376 | 467174 | 3717978 | 39.71 |
| 55-59 | 1128 | 4 | 0.00355 | 0.01757 | 0.98243 | 93247 | 1639 | 462138 | 3250803 | 34.86 |
| 60-64 | 889 | 5 | 0.00562 | 0.02773 | 0.97227 | 91608 | 2540 | 451690 | 2788665 | 30.44 |
| 65-69 | 862 | 5 | 0.0058 | 0.02859 | 0.97141 | 89068 | 2546 | 438973 | 2336976 | 26.24 |
| 70-74 | 756 | 7 | 0.00926 | 0.04525 | 0.95475 | 86521 | 3915 | 422820 | 1898003 | 21.94 |
| 75-79 | 497 | 10 | 0.02012 | 0.09579 | 0.90421 | 82606 | 7912 | 393251 | 1475183 | 17.86 |
| 80 y más | 478 | 33 | 0.06904 | 1 | 0 | 74694 | 74694 | 1081931 | 1081931 | 14.48 |

Para San Francisco entre 2001-2010, en el cálculo de los años potenciales de vida perdido por sexo se observó que para el año 2003 se presentó la mayor tasa de APVP para los hombres (APVP: 3 875; Tasa: 1 114.3 por mil habitantes), la mayor tasa de APVP para las mujeres se presentó en el 2001 (APVP: 624; Tasa: 193 por mil habitantes) (Tabla 5). Al observar los APVP por grupo de causa de defunción, se evidencio que para las causas externas y para las enfermedades del sistema circulatorio se presentaron mayor número de años potenciales de vida perdidos para este decenio, 12 028 y 575 respectivamente (Tabla 6), todas las demás causas, ciertas afecciones originadas en el periodo perinatal, neoplasias, enfermedades transmisibles y los síntomas, signos y afecciones mal definidas, presentaron 505, 423, 373, 369 y 288 número de APVP respectivamente.

Tabla 5. Años potenciales de vida perdidos por sexo y año de defunción, municipio de San Francisco – Antioquia, decenio 2001- 2010.

|  | Mujeres | | Hombres | |
|---|---|---|---|---|
|  | APVP | Tasa por 1000 hab. | APVP | Tasa por 1000 hab. |
| 2001 | 624 | 193.2 | 1507 | 415.2 |
| 2002 | 232 | 72.5 | 995 | 280.0 |
| 2003 | 482 | 152.8 | 3875 | 1114.3 |
| 2004 | 260 | 83.3 | 2478 | 729.7 |
| 2005 | 175 | 56.7 | 1193 | 360.5 |
| 2006 | 40 | 13.3 | 588 | 180.5 |
| 2007 | 213 | 71.7 | 655 | 204.8 |
| 2008 | 58 | 19.7 | 258 | 82.1 |
| 2009 | 305 | 106.5 | 228 | 74.0 |
| 2010 | 5 | 1.8 | 393 | 130.6 |

Tabla 6. Años potenciales de vida perdidos por causa de defunción agrupada con base en la lista 6/67 de la OPS (CIE-10), municipio de San Francisco – Antioquia, decenio 2001-2010.

| | Causa agrupada con base en la Lista 6/67 de la OPS (CIE-10) | | | | | | |
|---|---|---|---|---|---|---|---|
| | Enfermedades transmisibles | Neoplasias (tumores) | Enfermedades del sistema circulatorio | Ciertas afecciones originada en el periodo perinatal | Causas externas | Todas las demás causas | Síntomas, signos y afecciones mal definidas |
| APVP | 369 | 373 | 575 | 423 | 12028 | 505 | 288 |

**Discusión**

En general, para municipio de San Francisco, el número de defunciones durante los años 2001 a 2010 se presentaron en mayor proporción para los hombres con un 75.7% (863 defunciones por cada 100,000 hombres) en comparación con las mujeres (306 muertes por cada 100,000 mujeres), las personas que estaban casadas presentaron una proporción de defunciones de 29.6% (185 muertes por cada 100,000 habitantes); se encontró que San Francisco presentó defunciones en mayor proporción en las personas con un nivel educativo de primaria con un 28.1% (175 defunciones por cada 100,000 habitantes); las personas del régimen subsidiado presentaron una mayor proporción de defunciones 48.1% (299 defunciones por cada 100,000 habitantes), de igual forma que el área rural disperso con un 56.7% (353 muertes por cada 100,000 habitantes). Apoyado en las anteriores tasas específicas de mortalidad es posible identificar la parte de la población que durante el decenio contribuye mayormente al número de defunciones y que pareciera ser que presentan mayor riesgo de sufrir un evento de muerte en comparación con los que registraron menores tasas de mortalidad.

Los enfrentamientos entre el ejército, las autodefensas y las guerrillas del ELN y las FARC desencadenaron temor y amenazas entre los habitantes de San Francisco, quienes se desplazaron entre los años 1998 y 2005, esta violencia fue el principal factor de causas de defunción para nuestro decenio de estudio. (18) Esto se hizo más evidente en la población masculina, en donde la proporción de defunciones por el grupo de causas de agresiones y homicidios, inclusive secuela fue de 94% (249 muertes por cada 100,000 habitantes) en comparación con las mujeres (16 muertes por cada 100,000 habitantes). Los jóvenes hombres fueron los que presentaron mayor número de defunciones durante el decenio, la mayor tasa de mortalidad (68 por cada 100,000 habitantes) se presentó en el grupo de edad de 20 a 24 años, marcando una gran diferencia con respecto a la de las mujeres donde la mortalidad fue más baja (3 por 100,000 habitantes). Estas altas tasas de mortalidad en la población joven masculina es también el reflejo de américa latina, en donde la violencia afecta a toda la población y primordialmente a los jóvenes hombres, trayendo como consecuencia altos costos económicos y sociales para la población, que podrían ascender a los 10.000 millones de dólares al año para el país (19).

Si para San Francisco, comparamos las tasas de mortalidad por causas externas (384 muertes por cada 100,000 habitantes) con las del departamento de Antioquia para el año 2007 con una tasa de 82 casos por cada 100,000 habitantes, podemos notar una diferencia de éstas, ya que se dan en el mismo decenio, y que, aunque no se encuentran estandarizadas para su comparación, nos da una idea de la magnitud de defunciones ocurridas en el municipio con respecto al departamento (19).

Se presentó en las tasas de mortalidad por sexo un pico de la línea de tendencia para los hombres en el grupo de edad 20-24 años, indicando una mayor mortalidad para ellos en este grupo de edad. Estas altas tasas de mortalidad en la población joven igualmente son preocupantes para el departamento de Antioquia, donde se presenta un incremento en la tasa de muertes por homicidios y por sexo en los grupos de edad de: 5 a 9 y 10 a 14 años. Este comportamiento de un alto porcentaje de muertes de hombres en edad productiva, representa un alto costo económico y social, que afecta para la población, en años de vida perdidos, costos de atención, sensación de inseguridad para toda la población, entre otros (19).

La historia de violencia que ha marcado el municipio de San Francisco se ha visto reflejada en los años potenciales de vida perdidos (APVP), los cuales fueron para las causas externas y para las enfermedades del sistema circulatorio 12 028 y 575 respectivamente, siendo estos dos grupos de causas los que predominaron durante el decenio. Se observó que en promedio los APVP en ese mismo periodo fueron mayores en los hombres en comparación con las mujeres, excepto en los años 2008, 2009 y 2010 donde los APVP en promedio fueron mayores para las mujeres. Es importante tener en cuenta que los años potenciales de vida perdidos son uno de los efectos más impactantes de las muertes por homicidio, que representaron para el departamento de Antioquia en el año 2006, 109 812 años potenciales de vida perdidos y en el año 2010 se perdieron 76 928 años más, la mayoría en hombres en edad productiva, produciendo grandes costos humanos, económicos y sociales (19). Esta situación que aqueja al municipio de San Francisco durante el decenio, no difiere de la situación departamental, y es importante tener en cuenta la implementación de políticas públicas que ayuden a disminuir los homicidios, y la violencia en general, ya que se ve reflejada en los resultados de los estudios de mortalidad, tanto para el municipio como para el departamento en general.

La esperanza de vida al nacer para los hombres durante el decenio fue de 65.55 años y para las mujeres fue de 85.38 años, que en comparación con la esperanza de vida al nacer para Colombia en el año 2010 (hombres 69.78 años y mujeres 77.14 años) (20) San francisco presenta en los hombres una diferencia de solo 4 años aproximadamente siendo esta menor para el municipio, contrario a las mujeres que presentan una diferencia de 8 años aproximadamente, siendo mayor para el municipio. La esperanza de vida es un importante indicador de los resultados que se obtienen al disminuir el nivel de mortalidad general de una población (21), para el municipio es importante disminuir las tasas de homicidio y agresiones, como causas externas, ya que es la que mayor afecta en la mortalidad de dicho municipio; con ello se reflejaría un aumento en la esperanza de vida, primordialmente en los hombres que presentan menor esperanza de vida en comparación a la nacional.

Las investigaciones en su desarrollo pueden emplear la observación o la intervención física, química o psicológica. De igual forma puede generar registros o, como es nuestro caso, utilizar datos existentes que contengan información biomédica y demás acerca de los individuos, identificados o no a partir de la misma (22).

Teniendo en cuenta estos aspectos y lo relacionado a los datos existentes, en esta investigación damos fe de que estos registros provienen de una fuente de información válida y de buena calidad, confiando en la veracidad de la información y la no falsificación de la misma. Se destacan las bases teóricas que se tiene para garantizar un buen diseño del estudio y como investigadores nos entregamos profesionalmente, con valores, responsabilidad y conocimiento para su desarrollo y publicación de dichos resultados. Igualmente, que los procedimientos ejecutados se ajusten a los objetivos y lograr a plenitud el alcance de los mismo (22).

En garantía de un estudio único y ético, como investigadores debemos velar por la seguridad de la información, por proteger su confiabilidad y restringir su acceso a terceros, de esta forma damos fe del valor que tiene la información para los investigadores y para la institución que permite el acceso y utilización de la misma.

Por ello velamos y trabajamos para presentar resultados investigativos que den información segura y verídica de la situación de la mortalidad registrada por causa y las características sociodemográficas, en el municipio de San Francisco - Antioquia, 2001-2010.

**Referencias**